\documentclass[12pt]{article}
\pdfoutput=1
\usepackage{epsfig}
\usepackage{graphicx}
\usepackage{cite}
\usepackage{amsfonts}
\usepackage{amssymb}
\usepackage{bm}
\usepackage{latexsym}
\usepackage{amsmath}
\numberwithin{equation}{section}
\def\ee{\end{equation}}
\def\ba{\begin{eqnarray}}
\def\ea{\end{eqnarray}}

\def\bq{\begin{quote}}
\def\eq{\end{quote}}

 at 10truept

\newcommand{\beq}{\begin{equation}}
\newcommand{\eeq}{\end{equation}}
\newcommand{\beqa}{\begin{eqnarray}}
\newcommand{\eeqa}{\end{eqnarray}}
\newcommand{\bea}{\begin{eqnarray}}
\newcommand{\eea}{\end{eqnarray}}
\newcommand{\p}{\partial}

 \newcommand{\ep}{\epsilon}

\newcommand{\vect}[1]{\bm{\mathrm{{#1}}}}

\newcommand{\Hotps}{\left(\frac{H}{2\pi}\right)^2}

\def\ltap{\ \raise.3ex\hbox{$<$\kern-.75em\lower1ex\hbox{$\sim$}}\ }
\def\gtap{\ \raise.3ex\hbox{$>$\kern-.75em\lower1ex\hbox{$\sim$}}\ }
\def\gl{\ \raise.5ex\hbox{$>$}\kern-.8em\lower.5ex\hbox{$<$}\ }
\def\roughly#1{\raise.3ex\hbox{$#1$\kern-.75em\lower1ex\hbox{$\sim$}}}

\def\kq{{\kappa_q}}
\def\kqi{{\kappa_{q,i}}}
\def\kqj{{\kappa_{q,j}}}
\def\kk{{\kappa_k}}

\begin{document}

\thispagestyle{empty}
\begin{titlepage}
\nopagebreak

\title{
 Cosmological observables, IR growth of fluctuations, and scale-dependent anisotropies} 
 

\vfill
\author{Steven B. Giddings$^{a}$\footnote{giddings@physics.ucsb.edu}\ \  and Martin S. Sloth$^{b}$\footnote{sloth@cern.ch}
}
\date{ }


\maketitle

\vskip 0.5cm

{\it  $^{a}$ Department of Physics, 
University of California, Santa Barbara, CA 93106}
\vskip 0.5cm
{\it  $^{b}$ CERN, Physics Department, Theory Unit, CH-1211 Geneva 23, Switzerland}

\vfill
\begin{abstract}We extend semiclassical methods in inflationary cosmology that capture leading IR corrections to correlators.  Such large IR effects can be absorbed into a coordinate change when examining sufficiently local observables, but not when comparing observations at large separation in scales, such as seen by a late-time observer.  The analysis is facilitated by definition of a scale-dependent metric and physical momentum.  These assist definition of   ``IR-safe" observables seen by a post-inflationary observer, which are contrasted to those based on the local geometry of the reheating surface.  For such observables, the observer's horizon provides an effective IR cutoff.  IR growth contributes to enhanced statistical inhomogeneities/anisotropies at short scales, observation of which by a present day observer might be sought in 21 cm measurements.  Such IR corrections are argued to grow large for a very late-time observer.

 \end{abstract}
 \vskip.4in
 
\noindent
CERN-PH-TH/2011-070\hfill \\  
\vfill
\end{titlepage}

\setcounter{equation}{0} \setcounter{footnote}{0}

\section{Introduction}

While inflation has successfully resolved a number of conceptual questions, and nicely matches existing cosmological data, it has raised some deeper puzzles.  Notable among these is the problem of large infrared (IR) effects.  The simplest of these arises from a very basic mechanism (see, {\it e.g.},\cite{Linde:2005ht}):  inflation produces fluctuations of a light field outside the horizon; these fluctuations then ``freeze" in amplitude, and the cosmological expansion causes them to accumulate at long wavelength.  This produces IR divergences in correlation functions, and is suggestive of a strongly fluctuating structure on the longest scales.

The spectrum of light fields depends on the model considered, but includes the inflaton, and certainly the modes of the graviton.  Indeed, one guise of this effect is self-reproducing inflation, in which inflaton fluctuations produce a large-scale spacetime structure with regions of very different effective cosmological constant.  An important question is how to define sensible observables in such a situation, particularly in light of the IR growth.  Specifically, the massless tensor graviton modes are always present, and are described by correlators that exhibit both IR divergences and IR growth at long times.  An important question regards whether these have observable consequence.

Part of the problem is how to formulate gauge-invariant observables, which is an outstanding issue in quantum gravity; diffeomorphism invariance implies that local observables familiar from field theory are not gauge invariant.  This indicates that observables must take a more nonlocal form.  On the practical level, one would like to understand formulation of such observables describing our actual observations; on a more formal level one would like to find, {\it e.g.}, gauge-invariant observables that reduce to the local observables of field theory in certain approximations\cite{GMH}.  

In fact, it seems useful to have language to distinguish these two notions.  
If we imagine that there is a yet-unknown complete quantum-mechanical description of cosmology, then one of its basic features should be a set of gauge-invariant quantum-mechanical observables; we will refer to these as {\it q-observables}.  These, however, may or may not be observable by us given our limitations as Earth-bound observers in a particular era of cosmology.  Thus there is a much more restricted set of in-practice {\it observables}.  Of course, an important question is what q-observables are actually observable.  One possible way to think of this is as requiring a {\it portal},\footnote{Analogous to a similar notion in describing hidden-sector physics.} which is a mechanism for the q-observable to imprint its information in  physics visible to us.

Returning to metric fluctuations, then, there are two questions: first, what are gauge-invariant q-observables, for example sensitive to the large IR fluctuations, and second, are there effects of these which are actually observable.  This paper will advance positive arguments on both fronts, and in the process address the question of how to formulate ``IR-safe" observables free of IR divergences.

Consider a perturbation of an inflating cosmological metric, with flat spatial sections and scale factor $a(t)$,
\beq
ds^2=-dt^2 +a^2(t)(e^{\Gamma})_{ij} dx^idx^j\ ,
\eeq
where $\Gamma_{ij}$ parameterizes the perturbation, and can have both scale and traceless tensor parts.  Such perturbations redshift to long wavelengths, and are at least locally unobservable.  Indeed, in the long-wavelength limit, $\Gamma_{ij}$ is a constant, and can be removed by a change of coordinates:
\beq\label{constxm}
{\tilde x}_i = (e^{\Gamma/2})_{ij}x_j\ .
\eeq
this is seen if one attempts to formulate observables in terms of local scalar quantities, such as the curvature; the curvature due to the perturbation vanishes in the limit of large redshift.

However, the perturbation is nonetheless still present in the geometry investigated at longer length scales, and one question is how to describe this presence in terms of suitable q-observables.  We will present aspects of this story in future work\cite{GS-toappear}, and instead focus on the more practical question -- can such strong fluctuations manifest themselves in quantities we actually observe?

To summarize, \cite{GS} outlined a story of this nature, understood via a semiclassical picture.  Specifically, scalar and tensor fluctuations can contribute to observable quantities such as the microwave background fluctuations via their gravitational couplings to shorter-wavelength fluctuations.  At short distances compared to the longer mode's wavelength, its effect can be removed by a transformation such as \eqref{constxm} and is thus invisible.  However, for the purposes of observations of a more global nature, the fluctuation persists, and alters the spectrum because of mismatches in the local metric at points separated by long distances, due to the effect of the long wavelength modes.  The effect on the spectrum can be understood by noting that the spectrum will be determined in terms of the local proper momentum when the corresponding perturbations leave the horizon scale.  By comparing spectra over a sufficiently large hierarchy of scales, the long-wavelength fluctuation can be detected, via its distortion of the spectrum.  This also indicates a resolution of the IR divergence problem:  we can transform away fluctuations that have wavelengths long as compared to the region we observe.  But -- a late-time observer sees more and more fluctuating volume, and so can see an IR growth of fluctuations, apparently leading to a breakdown of such a perturbative treatment.

The present paper will further develop and clarify this picture.  The next section outlines a specific example framework, single-field slow-roll inflation, although the conclusions extend more broadly.  The third section introduces the notion of a scale-dependent metric perturbation and proper momentum, and uses these to give a simple description of the effect on the spectrum of fluctuations, extending \cite{GS}.  The fourth section describes large-scale effects of metric fluctuations, contrasting their description in terms of the geometry of the reheating surface, to that in terms of the observations of late-time observers such as ourselves.  In the process, we also give a prescription that eliminates IR-cutoff-dependent effects due to presently-unobservable longer wavelength fluctuations. But, tensor fluctuations can lead to statistical inhomogeneities/anisotropies that one might try to observe in 21 cm measurements, by comparing scalar fluctuations at different observable locations and scales.  We conclude with general comments on relation to renormalization group ideas, and to deeper questions in quantum gravity.

\section{Action and perturbation expansion}
\label{sec2}

For a simple example framework, consider single-field, slow-roll inflation, with inflaton field $\phi$ and potential $V$:
\beq\label{action}
S=\frac{1}{2}\int\sqrt{-g}\left[{R\over 8\pi G}-\p_{\mu}\phi\p^{\mu}\phi -2V(\phi)\right]~.
\eeq
$R$ is the Ricci scalar, and we choose units $8\pi G=1$.

A perturbative description of the coupled metric and matter fluctuations can be derived
using the Arnowitt--Deser--Misner\cite{Arnowitt:1960es} (ADM) parameterization of the line element,
\begin{equation}\label{admmetric}
    ds^2= -N^2 dt^2 + h_{ij}(dx^i + N^idt)(dx^j + N^jdt)~,
\end{equation}
where $N$, $N^i$ are the lapse and
the shift functions. 
We further decompose the metric as
\beq
h_{ij}= a^2(t) e^{2\zeta(x,t)} [e^\gamma]_{ij}\ ,
\eeq
where $a(t)$ and $\phi=\phi_0(t)$ parameterize the classical, homogeneous slow-roll solution, and $\gamma_{ii}\equiv \delta^{ij}\gamma_{ij}=0$.  The physical tensor degrees of freedom are contained in $\gamma_{ij}(x,t)$, and a scalar degree of freedom in a combination of $\zeta$ and $\phi$ depending on gauge.  

The dynamics are found by expanding \eqref{action} in $\zeta$, $\gamma_{ij}$, and $\phi-\phi_0$; the lapse and shift are determined by their equations of motion, the constraints.  The quadratic action determines a mode decomposition for the degrees of freedom.  For example, in the gauge $\partial_i\gamma_{ij}=0$, 
\begin{equation}\label{gravdecomp}
     \gamma_{ij}(x) = \sum_{s=+,\times} \int
    \frac{d^3 k}{(2\pi)^{3}}\left[
        b^{s}_{\vect{k}}\ep^s_{ij}(\vect{k})\gamma_k(t)
        + b^{s\dagger }_{-\vect{k}}\ep_{ij}^{s*}(-\vect{k})\gamma^{*}_k(t)\right]
        e^{i\vect{k}\cdot\vect{x}}=\int
    \frac{d^3 k}{(2\pi)^{3}}\gamma_{ij}(k;t,x) ~,
\end{equation}
where we defined the modes $\gamma_{ij}(k;t,x)$ for later use, and
 $b^{s}_{\vect{k}}$ is an annihilation operator, corresponding to helicity $s=+,\times$, and satisfying
 \beq
\left[ b^{s}_{\vect{k}},  b^{s\dagger }_{\vect{k}'}\right] =(2\pi)^3  \delta_{ss'}\delta^3\left(\vect{k}-\vect{k'}\right)\ .
\eeq
The polarization tensors
$\ep^s_{ij}$ are chosen to satisfy
transversality and tracelessness conditions, along with
the completeness relation $\ep_{ij}^s(\vect{k})\ep_{ij}^{\ast s'}(\vect{k})=2\delta_{ss'}$.  The mode functions $\gamma_k(t)$ depend on the slow-roll potential and solution $\phi_0$.

The two-point function gives an important measure of the gravitational fluctuations; specifically, consider the double trace,
 \beq\label{2pt}
\langle \gamma^2(x,x')\rangle = \frac{1}{4}\langle \gamma_{ij}(x) \gamma_{ij}(x')\rangle =\int{d^3k\over (2\pi)^3}|\gamma_k(t)|^2 e^{i\vect{k}\cdot(\vect{x}-\vect{x'})}\  .
\eeq
The coincident limit $\vect{x}=\vect{x'}$, which gives the variance, is ultraviolet (UV) divergent, but may be regulated by choosing a minimum physical separation $a(t)|\vect{x}-\vect{x'}|\sim 1/H$, at the Hubble scale $H={\dot a}/a$, effectively providing a UV cutoff at $k\approx  a(t)H$.  Eq.~\eqref{2pt} can also be large in the IR, for inflation of sufficient duration.  For example, in the case of de Sitter space, $V=\Lambda=$const., the mode functions are easily written in terms of the conformal time $\eta=\int dt/a(t)=-1/(Ha)$, 
\beq\label{modefcns}
 \gamma_k (\eta) = 
    \frac{H}{\sqrt{k^3}}(1+ik\eta)e^{-ik\eta}\ .
 \eeq
The variance \eqref{2pt} then has a logarithmic IR divergence.  We can artificially introduce a comoving IR cutoff $L$, which we might imagine as parameterizing a finite beginning of inflation, in which case
\beq\label{vargrowth}
\langle \gamma^2(x)\rangle = 2\Hotps\int_{1/L}^{ a(t)H} \frac{dk}{k}(1+k^2\eta^2)\approx 2\Hotps \log[a(t)HL]\ ,
\eeq

There can be similar growth in slow-roll, and also for the variance of the scalar mode.  
One current goal will be to improve understanding of the physical implications of this growth, and of the physical framework necessary to eliminate the artificial dependence on the IR cutoff $L$.

\section{Contributions of IR modes: semiclassical relations}

We will be interested in the effects of buildup of IR modes on cosmological observables such as the scalar spectrum.  These can be calculated in various frameworks and gauges; for our discussion we will fix the time slicing during inflation by working on slices of constant inflaton,
\beq\label{comgauge}
\phi=\phi_0(t)\ .
\eeq
This comoving condition leaves gauge freedom corresponding to choice of spatial coordinates.  This can be further fixed via the transversality condition, $\partial_i\gamma_{ij}=0$.  
One can then show that in this gauge, the constraints are consistent with $N-1$ and $N^i$ vanishing at first order in the perturbations.

The two point functions for $\zeta$ and $\gamma$ are then computed at gaussian (free-field) level by standard methods, as described above.  Our current focus is on one loop corrections to these correlators.  These were computed by two methods in \cite{GS} for de Sitter space:  a direct one-loop calculation, and via the semiclassical relations outlined there.  These were found to agree in some detail, and it was argued (with additional supporting evidence) that the latter method also applies to slow-roll. Further elaboration and checks were given in \cite{Byrnesetal,Xue:2011hm,ChMa,Riotto:2011sf}.

Specifically, in the latter semiclassical method, the correlators are obtained from a simple and intuitive physical picture:  when a given mode exits the horizon, the spectrum is determined in terms of the {\it physical} momentum, computed by treating the longer-wavelength modes as providing a background metric.

In order to formulate this condition more precisely, we will define a notion of a {\it scale-dependent metric fluctuation}, $\Gamma_{ij}(q,t,x)$ at scale $q$, via the formula
\beq\label{gammadefa}
\Gamma_{ij}(q,t,x) = \int^q_{L^{-1}} {d^3k\over (2\pi)^3} \left[ 2 \zeta(k; t,x)\delta_{ij} + \gamma_{ij}(k; t,x)\right]\ 
\eeq
where $\zeta(k; t,x)$ and $\gamma_{ij}(k; t,x)$ are mode functions with comoving momentum $k$, as in the expansion \eqref{gravdecomp}.  In particular, if we restrict to the domain $q<a(t)H(t)$, this expression will be constant in time to a very good approximation, as excitations are frozen outside the horizon, but will vary at comoving distances $\gtrsim 1/q$.
One can then define a {\it scale-dependent physical momentum} (modulo rescaling by $a(t)$) via
\beq\label{scaledepp}
\kqi(k,x) = [e^{-\Gamma(q,x)/2}]_{ij} k_j \quad ; \quad \kq^2(k,x) = [e^{-\Gamma(q,x)}]_{ij} k_i k_j\ ,
\eeq
as well as a scale-dependent metric, $h_{ij}(q; x,t)=a^2(t) [\exp\{\Gamma(q,x)\}]_{ij}$.  These expressions are again constant in $x$ in fixed regions of comoving size $\ll 1/q$, but vary on scales $\gtrsim 1/q$. An extended statement of the proposal of \cite{GS} is   that  leading IR-dependent higher-order effects are incorporated into the spectrum $P$ by writing the {\it tree-level} two-point function $P_0(k)$ instead as a function of $\kk(k,x)$: 
\beq\label{resum}
P(k,x)d^3k = P_0(\kk(k,x)){d^3\kk }\ .
\eeq
That is, the physics determines the spectrum in terms of the instantaneous {\it physical} momentum of the mode when it is exiting the horizon, as expected. 
The same reasoning applies whether the fluctuation in question is in $\zeta$, the tensor modes $\gamma_{ij}$, or in another ``spectator" field $\sigma$. For example in the former case, we have the definitions\footnote{In a similar definition for $P$, the delta function is effectively smeared on scales $q$.}
\beq
\langle \zeta _{{\vect k}} \zeta _{{\vect k}'}\rangle_0 = (2\pi)^3 \delta^3({\vect k}+{\vect k}') P_{\zeta,0}(k)\quad ,\quad P_{\zeta,0}(k) = {(2\pi^2) {\cal P}_{\zeta,0}(k)/ k^3} 
\eeq
and can parameterize ${\cal P}_{\zeta,0}(k)\propto k^{n_s-1}$, with $n_s=1$ the scale-independent case.

 Note that it is simply the Taylor expansion of the right hand side of \eqref{resum} around the {\it tree-level} two-point function $P_0(k)$ that gave us the order-by-order corrections to the power spectrum in \cite{GS}, as can be seen from {\it e.g.} eq. (4.5) or eq. (4.11) of that paper.  The prescription \eqref{resum} extends to higher-point functions as well, with examples given in \cite{GS}.

As a function of $k$, $P(k,x)$ is sensitive to the IR cutoff, which enters via \eqref{gammadefa}.  This is what explicit loop calculations show, in de Sitter space\cite{GS}.  However, rewriting $P$ as a function of $\kk(k,x)$ resums leading IR logs, giving a candidate ``IR-safe" quantity $P_0(\kk(k,x))d^3\kk$, as is shown by the IR match between the expression \eqref{resum} and the one-loop calculation.  This latter aspect has been emphasized in other recent discussions \cite{Byrnesetal,Senatore,Hebecker2}. Ref.~\cite{Urakawa:2010kr} has also demonstrated cancellation of IR tensor variance contributions, in a position-space analog of 
\eqref{resum}.  We will give an IR-safe prescription for  actual observers in section 4.

\section{Late-time observations and IR growth}

An important question is how these corrections contribute to the spectrum of fluctuations seen by a late time observer, such as us; this is also a sharper context to investigate IR safety.  We first consider the reheating surface, at $\phi=\phi_r$ which is a constant time slice, $t=t_r$, in our gauge \eqref{comgauge}.  Fluctuations with a longer wavelength than the horizon scale at that time will have been generated with the spectrum $P$ in  \eqref{resum}, and then stay constant outside the horizon.  Thus, for $\kk(k,x)<a(t_r)H(t_r)$, these fluctuations will be given by this spectrum.

\subsection{Geometry(ies) of the reheating surface vs. late time observers}

One possible approach, also advocated in \cite{Senatore,Urakawa:2010kr,Hebecker2,Tanaka:2011aj}, is to describe the fluctuations, geometrical or otherwise, in terms of the geometry of  the reheating surface.
This geometry is determined by the spatial metric(s) $h_{ij}(q,x,t_r)$.  This will not, in general, be flat, but can be made ``as flat as possible" at scale $q$ by a coordinate transformation of the form ({\it c.f.} \eqref{constxm})
\beq\label{flatcoord}
{\tilde x}^q_i = [e^{\Gamma(q,x)/2}]_{ij} x_j\ .
\eeq
For example, if one takes $q=a(t_r)H(t_r)$, incorporating all fluctuations outside the horizon,\footnote{For improved higher-order accuracy, this $q$ can be replaced by that determined by $\kq(q,x)=a(t_r)H(t_r)$.} and with the common assumption that fluctuations inside the horizon are unimportant, this puts the metric in a locally flat form at the (arbitrarily-chosen) origin of coordinates.  However, this will then introduce gradient terms in the metric of the form $\partial_i \Gamma_{jl}(q,x) x^l$, which become important at sufficiently long distances.  One has put the fluctuation spectrum \eqref{resum} in an apparently simpler form, since the conjugate variable to ${\tilde x}^q_i$ is \eqref{scaledepp},
up to the same kind of gradient terms.  Thus, in the vicinity of the given scale, the spectrum can be written as the uncorrected spectrum $P_0(\kq)$.  

In comparing the spectra at significantly different scales on the reheating surface, there will however be mismatches depending on which choice of scale-dependent metric is used in the definition \eqref{flatcoord}.  
Moreover, there have been suggestions\cite{Senatore,Urakawa:2010kr,Hebecker2} that gauge-invariant observables are naturally formulated in terms of correlators at fixed proper distance on the reheating surface.\footnote{One can in particular write the spectrum $P$ as a function of the average physical momentum on the reheating surface, $\langle \kappa_{q_r}(k,x)\rangle$, computed at the reheating horizon scale $q_r= a_rH_r$.  Written in terms of this variable, one finds a different apparent shift in the spectrum, of the form $(n_s-1) ( \langle \zeta^2\rangle_r - \langle \zeta^2\rangle_{h.c}) P_0/2$, where h.c. denotes horizon crossing; this is like that found in \cite{Senatore}.} For distances much smaller than the inverse $1/q$ of the scale used in defining the scale-dependent metric \eqref{gammadefa}, its geodesics are those of the unperturbed metric, corresponding to the fact that eq.~\eqref{flatcoord} is effectively just a constant transformation.  But, at longer distances than $1/q$, the fluctuations modify the geodesics, so the scale used to define the metric is relevant.  Further examination of this and q-observables characterizing metric fluctuations in the geometry of the reheating surface will be given in \cite{GS-toappear}.

However, a late-time physical observer does not directly measure the local geometry of the reheating surface.\footnote{Indeed, if we made observations in terms of proper momentum at reheating, our observations would depend on the details of reheating, while the  point of working in terms of conserved correlations at horizon exit is that CMB predictions do not depend on such microphysics.} In fact, at the time of reheating, the perturbations on  scales relevant for observations today are far outside the horizon, and when comparing correlation functions as observed today, corresponding to different causally-disconnected regions on the reheating surface, one has to take into account that the local physical momenta in different regions are different, via \eqref{scaledepp}, due to intermediate wavelength fluctuations present within the observable universe at the reheating time.

To a late time observer like us, it is the physical momentum {\it today} that matters, and today the intermediate wavelength fluctuations, which were background for shorter-wavelength fluctuations at the time of reheating, have entered the horizon and can no longer be counted as background. Thus, for an observer today, observing a largest mode entering the horizon, which exited the horizon $60$ e-folds before the end of inflation, the relevant physical momentum coincides with the physical momentum when that mode exited the horizon.  This defines a notion of a ``small-box" and the corresponding comoving momentum scale. For a  later observer, a  larger box is relevant.

\subsection{Late-time observation}

We therefore focus on the spectrum seen by  a late-time observer.  This is due to an essentially familiar effect:  at recombination, the fluctuations in $\zeta$ are imprinted in the observed $\Delta T/T$ of what are now microwave photons.  This follows from the well-developed treatment of the Sachs-Wolfe effect on large angular scales and the acoustic oscillations on smaller scales, which can be conducted precisely by following  the metric \eqref{admmetric} into the matter-dominated regime.  The metric perturbations governed by \eqref{resum} then reenter the horizon at times $t_p$ when $\kk(k)=a(t_p) H(t_p)$, and begin oscillating and decaying.  However, the leading effect of all this physics is simply the conversion of $\zeta(x,t_r)$, given in the {\it comoving} coordinates, into $\Delta T(\theta, \phi)/T$. Of course $\Delta T(\theta, \phi)/T$ can be written gauge invariantly and one is allowed to compute it any coordinates one likes. However, in a standard treatment, like the calculation of the second order Sachs-Wolfe effect in \cite{Boubekeur:2009uk}, one is simply calculating $\Delta T(\theta, \phi)/T$ in terms of the primordial curvature fluctuations in comoving coordinates. Thus, in this approach any new primordial second order effects are compared to the primordial linear spectrum in comoving coordinates.    

The physics we are looking at  is a small second order effect on the primordial spectrum of comoving curvature perturbations exhibiting mode-mode coupling of long wavelength modes with short wavelength modes. It can easily be summarized in the following way. Today we observe the last-scattering region, which consists of many causally separate patches. When we observe a correlation function of relative short fluctuations in many different places on the last-scattering surface, the background on which this correlation function is evaluated will change from place to place due to long wavelength fluctuations on the last-scattering surface. Of course only wavelengths larger than the wavelength of the correlation function we are evaluating, but smaller than the observed universe (i.e. the observed part of the last-scattering surface) will lead to any additional anisotropy. Wavelengths much larger than the observed universe today will only have an effect which can be absorbed in a coordinate transform \eqref{flatcoord}.

Thus, the spectrum of the needed $\zeta(x,t_r)$ is for the most part given by a function of the form \eqref{resum}, with $k$ conjugate to comoving coordinate $x$.  But, there is the important subtlety that  the metric perturbation $\Gamma_{ij}(q\sim a(t_0)H(t_0))$, where $t_0$ is the observation time, has not yet decayed. It is  moreover  apparently IR cutoff sensitive, or divergent.  However, at small scales, its effect can be eliminated by a transformation of the form \eqref{flatcoord}, with 
\beq
q_0=a(t_0) H(t_0)\ ,
\eeq
 (which again can be improved as in footnote three).  Specifically, at short  distances as compared to $H^{-1}(t_0)$, the gradient terms described above have effect parametrically small in  distance.  The transformation \eqref{flatcoord} simply corresponds to scaling out metric fluctuations frozen at longer scales, which are constant on the observer's Hubble scale.

The net effect is that we work in physical coordinates ${\tilde x}_{0,i}\equiv{\tilde x}^{q_0}_i$ in which the observer's horizon-scale metric appears flat.  And, we correspondingly rewrite the fluctuation spectrum \eqref{resum} in terms of the conjugate physical momentum, $p_{0,i}=[\exp\{-\Gamma(q_0,x_0)/2\}]_{ij} k_j/a(t_0)$, combining the relations to find
\beq\label{physmom}
\kk(k,x)^2 =a_0^2 [e^{\Gamma(q_0,t_0,x_0)-\Gamma(a_0p_0,t_r,x)}]_{ij} p_{0,i} p_{0,j}\ ,
\eeq
(modulo commutators) and the resulting spectrum given by \eqref{resum},
\beq\label{spec}
P(p_0)d^3p_0 = P_0\left[a_0\left([e^{\Gamma(q_0,t_0,x_0)-\Gamma(a_0p_0,t_r,x)}]_{ij} p_{0,i} p_{0,j}\right)^{1/2}\right]d^3\kk\ ,
\eeq
 which is now a function of the observed physical momentum $p_0$ and the comoving momentum $q_0$ corresponding the size of the observers horizon.  (Again, see footnote three.) 

Notice that this has accomplished something very important.  As a function of $k$, the spectrum depends on the IR cutoff $L$, and diverges as it is taken to infinity.  But, by working in terms of the observer's physical momentum $p_0$, where longer-wavelength fluctuations are ``scaled out," the observed spectrum instead depends on the quantity
\beq\label{gammaC}
\Gamma_{0,ij}(a_0p_0,t_0,x) = \Gamma_{ij}(a_0p_0,t_r,x)-\Gamma_{ij}(q_0,t_0,x_0)
\eeq
which is {\it IR safe} -- IR cutoff dependence is eliminated, and the observer's horizon size $1/q_0$ instead functions as an IR cutoff.

This is not to say that there are no IR large effects.  For sufficiently large hierarchy between the scale $p_0$ of the fluctuation being observed, and the horizon scale $H(t_0)$, the integral \eqref{gammaC} can make a large contribution.  Specifically,   looking at the tensor contribution, the variance, given in de Sitter space by eq.~\eqref{vargrowth}, signals such large contributions.  In this simple case  (which approximates slow-roll), we see that this variance grows linearly as
\beq\label{obsvar}
\langle\gamma^2\rangle \approx 2\Hotps \log[H(t_0)/p_0]\ .
\eeq
Thus, while this is small today, it becomes of order one for a late time observer who is able to see $1/H^{2}\sim S$ e-folds of inflation.\footnote{Here, in describing such a late-time observer, we neglect the effects of late-time inflation that appears to be beginning in the Universe today and interferes with such idealized observations.  Alternatively, if the present-day vacuum energy decays, ultimately the relevant information enters the observer's horizon.}   A more precise criterion is given in \cite{GS}, which examines the size of corrections in simple scenarios.  This appears to confirm the proposal of \cite{GS}, that large IR effects can be resummed/absorbed for the purposes of describing sufficiently local observables, but not for purposes of describing sufficiently global observables.  This also extends the discussion of \cite{Lyth,Bartolo:2007ti}, in which IR effects were argued to be small in the ``small box" but not in a ``large box."

The spectral distortion due to \eqref{gammaC} can be examined by expanding \eqref{spec}  in $\Gamma_{0,ij}$,
\bea\label{defspec}  
P_\zeta(p_0,x) = \Biggl\{P_{\zeta0}&+&\left[-\Gamma_{0,ij}(k,x)+ {1\over 2}\Gamma_{0,il}(k,x)\Gamma_{0,lj}(k,x)\right]k_i k_j {\partial P_{\zeta0}\over \partial k^2}\cr &+&{1\over 2}[\Gamma_{0,ij}(k,x)k_ik_j]^2 \left({\partial\over \partial k^2}\right)^2 P_{\zeta0}+\cdots\Biggr\}_{\Big|_{k=a_0p_0} } a_0^3e^{-Tr \Gamma_0(a_0p_0,x)/2}\ . 
\eea  
(See eqs. (4.5) and (4.11) of \cite{GS}; other formulas there can likewise be rewritten in terms of $p_0$.)  
The expectation value of the first-order term vanishes at leading order in perturbation theory.  However, by comparing different regions of the sky, the observer can in principle see differential isometries. In this way long wavelength modes induces a {\it statistical inhomogeneity/anisotropy} for correlation functions of shorter wavelength modes.  The size of the effects of the scalar and tensor modes   are proportional to the typical fluctuation in $\Gamma_{0,ij}$, determined by the accumulated variance \eqref{obsvar}, and thus is largest for the smallest-scale fluctuations. If we are measuring a correlation on a  comoving scale $a_0p_0$, the size of the effect of long wavelength tensor modes can be estimated to be of order the square-root of the variance of long wavelength tensor modes within the horizon, {\it i.e.}, from \eqref{obsvar},    
$\left(\langle \gamma^2\rangle_{obs} \right)^{1/2}\approx 2\times 10^{-5}$,
where as  example values we assumed a tensor-to-scalar ratio of order $r\sim 0.1$, the scalar amplitude as measured by WMAP7 $A_s = 2.46\times 10^{-9}$ \cite{Larson:2010gs}, and $p_0 = 10^3 H_0$. Planck is only expected to be able to probe statistical anisotropy down to a precision around $2\%$, but it has been predicted that it is possible to see statistical anisotropy all the way down to the $10^{-7}$ level by using  21 cm emissions to probe the ``dark ages," thereby improving statistics \cite{Pullen:2007tu}.  However, these statements apply to {\it homogeneous} anisotropies.  It would be interesting to investigate observability of the kind of inhomogeneous anisotropies discussed in this paper, in these futuristic 21 cm measurements.  
It is also interesting to note that while the present effect of tensor modes is imprinted on scalar modes of all scales, the effect of primordial tensor modes on $B$-mode polarization in CMB is only on relatively large scales.   This suggests that this effect could even be the most sensitive probe of primordial tensor modes in the far future.
Indeed, in a related work, Masui and Pen\cite{MaPe} gave arguments for 21 cm observability of the effect \eqref{defspec} on the spectrum, with a heuristic  re-derivation of the leading effect \cite{GS} based on similar considerations, but basing a signal to noise estimate on the approximation of a statistically homogeneous effect.

\section{Comments and conclusions}

This discussion is expected to generalize in several directions.  First, as initially investigated in \cite{GS}, one can apply a prescription like \eqref{spec} to higher-point functions, 
using  \eqref{physmom}. Secondly, while we have discussed the essential points in the context of single-field slow-roll inflation, clearly they have broader applicability.  For example, to control the statement that fluctuations exiting the horizon are determined in a simple way in terms of a physical momentum defined by the background, really all one needs is an adiabaticity condition that time scales for deviation from de Sitter space are long as compared to the Hubble time $1/H$,  here provided by slow-roll.   Likewise, the use of a scale-dependent metric and physical momentum have broader possible generality to other inflationary scenarios, and suggest an approach to defining certain IR-safe observables in dS space.

These notions also appear connected to renormalization group methods, of which they have a strong flavor, and which have been applied to inflation in related contexts\cite{Strominger:2001gp,Podolsky:2008qq,Burgess:2010dd}.
For example, thinking of $q$ as playing a role like a renormalization scale, one can differentiate to find renormalization group equations.  Applied to the scale-dependent momentum, this yields
\beq\label{pbeta}
q{\partial\over \partial q} \kqi(x) = -{1\over 2}\left({q\over 2\pi}\right)^3 \int d^2\Omega_q \left[ 2 \zeta({\vect q}; t,x)\delta_{ij} + \gamma_{ij}({\vect q}; t,x)\right] \kqj(x)\ 
\eeq
with related equations for other quantities.   One might think of the right side of \eqref{pbeta} as giving an analogue of a beta function.  Likewise, differentiating the spectrum \eqref{spec} gives
\beq
q{\partial\over \partial q} P = \left[q\partial_q\Gamma_{ij}(q,x)p_{0,i}p_{0,j}{\partial\over \partial p_0^2} +{1\over 2} q\partial_qTr \Gamma(q,x)\right] P\ ,
\eeq
a ``cosmological renormalization group" equation.  Note that IR cutoff dependence is also eliminated in these equations, and that similar equations can be written for other correlators.\footnote{Note that another alternative is to write such equations in terms of the more physical scale given by the variable $\kappa = \kq(q,x)$, instead of q.}

Finally, growth of the variance as in {\it e.g.} \eqref{vargrowth} means that for the late time observer seeing $S\sim 1/H^2$ e-folds, the one-loop corrections compete with the lowest-order effects.  This suggests a breakdown of this perturbative approach to gravitational fluctuations on these timescales, such as has been argued in \cite{Arkani,QBHB,Arkanietal,GS} to occur more generally on time scales $t\sim R S$, with here $R\sim 1/H$,  and supports the suggestion\cite{GS} that de Sitter space has a sort of instability, arising from accumulation of large-scale fluctuations.  Refs.~\cite{QBHB,GS} also argue this is parallel to a breakdown of the calculation of the quantum state of a black hole, also on the time scale $t\sim RS$, with $R$ the horizon radius, and $S$ the entropy (\cite{Arkani,Arkanietal} argue for a relation via an apparently different mechanism).
 Thus, there appear to be q-observables sensitive to the perturbative breakdown, describing late-time observation, and we have also argued that there is a portal to present-day observations where one could possibly see a small imprint of statistical anisotropy/inhomogeneity in short wavelength primordial correlation functions in 21 cm measurements.
These hint at a possible link between a potentially observable effect, and profound aspects of non-perturbative quantum gravity.

\vskip.1in
\noindent{\bf Acknowledgements} We wish to thank A. Hebecker, J. Lesgourgues, D. Marolf,   A. Riotto, and L. Senatore  for discussions. MSS would also like to thank the Physics Department at UCSB for kind hospitality when parts of this work was carried out. The work of SBG was supported in
part by the U.S. Dept. of Energy under Contract
DE-FG02-91ER40618.

\end{document}